\documentclass[pre,floatfix,notitlepage,nofootinbib]{revtex4-2}

\usepackage[utf8]{inputenc}
\usepackage{amsmath}
\usepackage{amssymb}
\usepackage{lipsum}
\usepackage{bm}
\usepackage{graphicx}
\usepackage{graphics}
\usepackage[dvipdf]{epsfig}
\usepackage{subcaption}
\usepackage{float}
\usepackage{wrapfig}
\usepackage{url}
\usepackage{epstopdf} 
\usepackage{xcolor}
\usepackage{tensor} 
\usepackage{setspace}

\newcommand{\bx}{{\mathbf x}}
\newcommand{\bX}{{\mathbf X}}
\newcommand{\bg}{{\mathbf g}}
\newcommand{\bG}{{\mathbf G}}
\newcommand{\bI}{{\mathbf I}}

\newcommand{\bEB}{{\mathbf{E}_{\text{B}}}}

\begin{document}
\title{Quadratic-stretch elasticity}

\author{E. Vitral and J. A. Hanna}
\affiliation{Department of Mechanical Engineering, University of Nevada,
    1664  N.  Virginia  St.  (0312),  Reno,  NV  89557-0312,  U.S.A.}

\begin{abstract}
A nonlinear small-strain elastic theory is constructed from a systematic expansion in Biot strains, truncated at quadratic order.
The primary motivation is the desire for a clean separation between stretching and bending energies for shells, which appears to arise only from reduction of a bulk energy of this type.
An approximation of isotropic invariants, bypassing the solution of a quartic equation or computation of tensor square roots, allows stretches, rotations, stresses, and balance laws to be written in terms of derivatives of position.
Two-field formulations are also presented.
Extensions to anisotropic theories are briefly discussed.
\end{abstract}
\date{\today}


\maketitle


\section{Introduction}

In a finite deformation, described in terms of positions or displacements, information about strains is polluted with information about rotations.  In statics, the strains are important but the rotations are irrelevant, and any hyperelastic energy and resulting constitutive relations must respect this.

A common solution is to symmetrize the deformation by ``squaring'' it with its transpose to obtain particular objective strain measures, either the right or left Cauchy-Green deformation tensors, the respective squares of the right and left stretch tensors.  This has the additional advantages that these measures, and the closely related Green-Lagrange or Euler-Almansi strain tensors, may be written straightforwardly using derivatives of position, and interpreted quite intuitively in terms of metric tensor components in referential and present configurations of the body.
Many other objective strain measures may be constructed as well, and a general nonlinear elastic energy can be written in terms of any of these.
In the continuum mechanics literature, this is often the end of the discussion.  One finds results derived for general stored energy functions, as well as a variety of phenomenological models of nonlinearly elastic materials, often expressed in terms of an incomplete collection of powers of principal stretches, for example Neo-Hookean, Mooney-Rivlin, or Ogden materials \cite{ogden1997nonlinear}.

However, in the physics literature, there is a desire to construct field theories by a systematic expansion in a small quantity. 
In linear elasticity, this would be the displacement, but in a nonlinear elasticity theory the choice of appropriate field variables is not immediately clear, and constitutes a fundamental unanswered question. 
Further motivation and clues for this question may be found in the study of thin shells, where it is often the case that strains are small but deformations are far from linear.  
Due to its ease of representation and interpretation, the Green-Lagrange strain has been widely adopted as the small field for expansion of elastic energies by soft condensed matter physicists \cite{Efrati09jmps, Dias11, Pezzulla15, NguyenSelinger17, vanRees17}.
However, the squaring of the deformation has the unfortunate consequence that the leading-order terms in such energies are quartic in stretch.
As has been rediscovered several times, surface elastic energies derived from dimensional reduction of such bulk elastic energies have undesirable features, including a mixing of stretching and bending contents \cite{
IrschikGerstmayr09, OshriDiamant17, Hanna19, WoodHanna19, OzendaVirga21}.
This stands in unsatisfying contrast to simple direct theories \cite{Antman68-2, Reissner72, WhitmanDeSilva74} that employ particular natural kinematic measures of surface stretching and bending to obtain separate and distinct energies.
 Selecting a simple bulk theory that corresponds to such simple direct theories requires a careful construction of energy quadratic in stretch, but this process has only been performed for one-dimensional and axisymmetric bodies \cite{IwakumaKuranishi84, Chaisomphob86, Magnusson01, IrschikGerstmayr09, OshriDiamant17, WoodHanna19}, and indeed appropriate bending measures have only been defined for such special cases\footnote{E. G. Virga has recently shared with us an unpublished note in which he derived more general bending measures.} of thin bodies \cite{Antman05, KnocheKierfeld11, OshriDiamant17, WoodHanna19}. 
Finally, quadratic-stretch elastic energies arise naturally in bead-spring descriptions of soft matter meso-structures such as biological membranes \cite{SeungNelson88, Deserno15}.

It is our intent in this paper to lay the groundwork for development of theories applicable to general shells, by first considering bulk elasticity from the perspective of a comprehensive expansion in stretches, truncated at quadratic order.
More precisely, we make use of the Biot strain, the deviation of the right stretch from the identity, as our small quantity.  As will become apparent below, this choice classifies our theory as ``referential'', but there is no reason why a complementary left-handed theory making use of tensors in the present configuration could not be constructed.
We also leave the analysis of reduced energies for thin bodies to subsequent work. 
A significant challenge in constructing a stretch-based theory is that stretch has an indirect dependence on position and its derivatives, requiring simultaneous consideration of an additional field, the rotation. 
Our synthesis leads us to connect several areas in theoretical and computational elasticity, including the kinematics of stretch and rotation, variational principles with auxiliary fields, and relations between isotropic invariants of different strain measures. 
We employ convected coordinates and a mixture of referential and present bases, revealing interesting relationships between components of various tensors.  Our Biot-quadratic energy is surprisingly rare in the nonlinear elasticity literature, although its roots go back to early work by Lurie \cite{lurie1968theory} and John \cite{john1960plane}.
Inspired by the works of Atluri and Murakawa \cite{atluri1977hybrid}, Wi{\'s}niewski \cite{wisniewski1998shell}, and Merlini \cite{merlini1997variational}, we propose a variational principle in terms of position, Biot strain, and rotation fields, and show that constitutive stress-strain relations and balance equations can be written in terms of positions and stretches alone. 
A new result is that the neglect of terms cubic in Biot strain allows for an explicit representation of stretch without either taking tensor square roots or solving a quartic equation, leading to an approximate description of all fields in terms of positions alone.

Our kinematic treatment may have broader application to theories in which rotations must be tracked, such as models of
nematic elastomers \cite{Modes11, Plucinsky18}
or tension-field theory \cite{Steigmann90}.
And although it is not the primary motivation for this work, we anticipate that it may have value with regard to the systematic construction of general nonlinear small-strain field theories of elasticity. 
However, one disadvantage of the present method as it currently stands is that it is not immediately clear how to adapt it to incompatible elasticity, in which no stress-free reference configuration or corresponding set of basis vectors exists but nonetheless the idea of a reference metric is still useful and intuitive.
We reserve this question for future work, and note that while the idea of an incompatible deformation gradient appears in previous works \cite{SadikYavari17, Nardinocchi13} it is not clear how such an object would be constructed in practice.

An outline of the paper follows. 
Section \ref{sec:kin} defines relevant tensorial objects and derives kinematic relations between them, their components, and their invariants.  Small-strain approximations lead to algebraic expressions for invariants and explicit expressions for all the relevant tensors in terms of derivatives of position.
Section \ref{sec:mec} presents stresses, balance laws, and constitutive relations obtained from a mixed variational principle in terms of position, Biot strain, and rotation.
In Section \ref{sec:energy}, a general isotropic energy quadratic in Biot strains is constructed.
Further constitutive relations are derived, in particular one for Biot stress in terms of Biot strain and its invariants, and then specialized to this energy.
Formulations are presented in terms of positions, positions and rotations, or positions and stretches.
Extensions to anisotropic theories are sketched in Section \ref{sec:ani}.
Appendices \ref{ap:dg}-\ref{ap:bell} discuss alternate decompositions of the deformation gradient, the symmetry of a certain rotated stress, details of the energy variation, and the strain and stress of Bell that complement those of Biot.

\section{Kinematics}
\label{sec:kin}

\subsection{Definitions and notation}
\label{sec:notation}

We denote material coordinates as $\eta^i$ and (noncovariant) material derivatives with respect to these as $d_i$.
An elastic body $\mathcal{B}$ with boundary $\partial\mathcal{B}$ has reference configuration $\bX(\eta^i)$ and present configuration $\bx(\eta^i)$, both in $\mathbb{E}^3$. 
We define referential and present coordinate bases $\bG_I = d_i\bX$ and $\bg_i = d_i\bx$, reciprocal bases through the relations ${\bf G}^I\cdot{\bf G}_J = \delta^I_J$ and ${\bf g}^i\cdot{\bf g}_j = \delta^i_j$ using the Kronecker delta, and covariant and contravariant components of the corresponding metric tensors $G_{IJ} = {\bf G}_I\cdot{\bf G}_J$, $g_{ij} = {\bf g}_i\cdot{\bf g}_j$, $G^{IJ} = {\bf G}^I\cdot{\bf G}^J$, $g^{ij} = {\bf g}^i\cdot{\bf g}^j$. Both metric tensors correspond to the identity, $\bG = \bg = \bI$.
In all of these expressions, capitalization of some indices is just a reminder that these should be raised and lowered with components of the reference metric; summation ignores case. 
We denote covariant derivatives corresponding to the referential and present metrics as $\bar\nabla_I$ and $\nabla_i$, respectively, and corresponding Christoffel symbols as $\bar\Gamma^{I}_{JK}$ and $\Gamma^i_{jk}$.
We define referential and present gradients $\bar\nabla() = \bar\nabla_i()\bG^I$ and $\nabla() = \nabla_i()\bg^i$.
The metric determinants $G = \mathrm{det} [G_{ij}]$ and $g = \mathrm{det} [g_{ij}]$, where $[\,]$ denotes a matrix, are found in the referential and present volume forms 
$dV = \sqrt{G} \prod\limits_{i} d\eta^i$ and $dv = \sqrt{g/G}\,dV$.
Surface forms $dA$ and $da$ are defined analogously using two surface coordinates only.

\subsection{Deformation measures, polar decomposition, and shifter}

The deformation gradient $\bar\nabla\bx$ takes the simple form $\bg_i\bG^I$ in material coordinates. 
While this is straightforwardly expressed purely in terms of derivatives of position, it is far from trivial to do the same for its rotationally-invariant stretching content.

Any non-singular second order tensor, such as the deformation gradient when $\sqrt{g/G} > 0$, admits unique right and left decompositions into an element of the special orthogonal group SO(3) of rotations and an element of the set Sym$^+$ of positive-definite symmetric tensors. 
In terms of the rotation tensor $\mathbf{Q} \in$ SO(3) and the right $\mathbf{U}$ or left $\mathbf{V}$ stretch tensors $\in$ Sym$^+$, 
\begin{equation}
    \mathbf{g}_i\mathbf{G}^I = \mathbf{Q}\cdot\mathbf{U} = \mathbf{V}\cdot\mathbf{Q} \, .
    \label{eq:polar}
\end{equation}
This is the most commonly adopted decomposition in continuum mechanics, but in Appendix~\ref{ap:dg} we consider alternatives.
The orthogonal rotation's transpose is also its inverse, 
\begin{equation}
	\mathbf{Q}^\top\cdot\mathbf{Q} = \mathbf{Q}\cdot\mathbf{Q}^\top = \mathbf{I} \, ,
	\label{eq:orthogonal}
\end{equation}
 a fact used repeatedly in the sequel.  
The right and left Cauchy-Green deformation tensors are 
$\mathbf{C} = \mathbf{G}^I\mathbf{g}_i\cdot\mathbf{g}_j\mathbf{G}^J = g_{ij}\,\mathbf{G}^I\mathbf{G}^J$ and $\mathbf{B} = \mathbf{g}_i\mathbf{G}^I\cdot\mathbf{G}^J\mathbf{g}_j = G^{IJ}\,\mathbf{g}_i\mathbf{g}_j$, respectively. 
Further relationships include $\mathbf{C} = \mathbf{U}^2$, $\mathbf{B} = \mathbf{V}^2$, 
$\mathbf{U} = \mathbf{Q}^\top\cdot\mathbf{V}\cdot\mathbf{Q}$, $\mathbf{V} = \mathbf{Q}\cdot\mathbf{U}\cdot\mathbf{Q}^\top$, $\mathbf{C} = \mathbf{Q}^\top\cdot\mathbf{B}\cdot\mathbf{Q}$, and $\mathbf{B} = \mathbf{Q}\cdot\mathbf{C}\cdot\mathbf{Q}^\top$.

A natural way to represent the rotation and stretch tensors is
\begin{equation}
    \mathbf{Q} = Q\indices{^i_J}\,\mathbf{g}_i\mathbf{G}^J \, ,\quad
    \mathbf{U} = U\indices{_I_J}\,\mathbf{G}^I\mathbf{G}^J \, ,\quad
    \mathbf{V} = V\indices{^i^j}\,\mathbf{g}_i\mathbf{g}_j \, ,
\label{eq:polar2}
\end{equation}
such that $\mathbf{Q}$ has the same mixed character as the deformation gradient, taking quantities in the reference configuration into the present configuration.  
From \eqref{eq:polar}, we find that the components of rotation, stretches, and metrics are related, 
\begin{equation}
	Q\indices{^i_J}G^{JK}U_{KL} = Q\indices{^i_J}U\indices{^J_L} = \delta^i_L = V\indices{^i_k}Q\indices{^k_L} = V^{ij}g_{jk}Q\indices{^k_L}\, .
    \label{eq:polar3}
\end{equation}
We further see that
\begin{equation}
	Q\indices{^i_J}g_{ik}Q\indices{^k_L} = G_{JL} \, , \quad Q\indices{^i_J}G^{JL}Q\indices{^k_L}= g^{ik} \, ,
    \label{eq:orthogonal2}
\end{equation}
which may be compared with
\begin{equation}
	U_{IJ}G^{IL}U_{KL}= g_{jk} \quad \mathrm{or} \quad U\indices{^I_J}U\indices{^J_K} = G^{IJ}g_{jk} \, .
    \label{eq:Usquared}
\end{equation}
Ericksen and Truesdell \cite{ericksen1957exact} describe this mixed-basis representation of rotation as ``shifted''. The components of the shifter \cite{malvern1969introduction} $\boldsymbol \mu = \left(\mathbf{g}^k\cdot\mathbf{G}_J\right) \mathbf{g}_k\mathbf{G}^J = \mu\indices{^k_J} \mathbf{g}_k\mathbf{G}^J$ can be used to relate the mixed-basis components of rotation with their representation in either the reference or present configuration.  The shifter can be thought of as the identity in a mixed-basis representation; the statement $\mathbf{Q}\cdot\boldsymbol\mu = \mathbf{Q} = \boldsymbol\mu\cdot\mathbf{Q}$ is, in components, $Q\indices{^i_k} \mu\indices{^k_J} = Q\indices{^i_J} = \mu\indices{^i_L}Q\indices{^L_J}$.  Invariants are computed along similar lines, for example 
$\textrm{Tr} \, \mathbf{Q} = \mathbf{Q}:\mathbf{I}
    = Q\indices{^i_J}\mathbf{g}_i\mathbf{G}^J:\mu\indices{^k_L}\mathbf{g}_k\mathbf{G}^L
    = Q\indices{^i_J}\mu\indices{_i^J}
    = Q\indices{_k^L}\mu\indices{^k_L}
    = Q\indices{^i_i} = Q\indices{_L^L}$.

\subsection{Stretch and rotation}

Due to the uniqueness of the square root of a symmetric positive-definite tensor \cite{stephenson1980uniqueness}, the right stretch $\mathbf{U}$ can be obtained as the square root of the right Cauchy-Green deformation $\sqrt{\mathbf{C}}\,$, or similarly, the left stretch $\mathbf{V}$ can be obtained as the square root of the left Cauchy-Green deformation $\sqrt{\mathbf{B}}\,$. 
However, this operation is inconvenient, and adds an additional layer of complexity, as these square roots, unlike $\mathbf{C}$ and $\mathbf{B}$ themselves, cannot be written in terms of derivatives of position. 
This has led many to explore other representations for $\mathbf{U}$ and $\mathbf{Q}$ that avoid the computation of tensor square roots, but instead make use of relationships between invariants of different strain tensors. 
While two-dimensional derivations such as that of Biot \cite{biot1965mechanics} are not too cumbersome, three-dimensional representations generally require another difficult step such as the selection of a root of a quartic equation \cite{hoger1984determination, Sawyers86, ting1985determination, stickforth1987square, lu1997direct, guan1998determination, jog2002explicit, bouby2005direct, norris2007invariants, scott2020u, sun2021explicit}.

Our starting point is Ting's relatively simple expression \cite{ting1985determination} for the right stretch $\mathbf{U}$ in terms of $\mathbf{C}$ and the invariants of $\mathbf{U}$. 
 The three principal invariants $i_k^{\textbf{U}}$ of $\mathbf{U}$ are
\begin{align}     \label{eq:inv}
    i_1^{\textbf{U}} &= \quad\;\; \textrm{Tr}\,\mathbf{U} \qquad\qquad\quad\;\;\; = \lambda_1 + \lambda_2 + \lambda_3 \, , \nonumber \\
    i_2^{\textbf{U}} &= \frac{1}{2}\Big[(\textrm{Tr}\,\mathbf{U})^2-\textrm{Tr}\,(\mathbf{U}^2)\Big]  = \lambda_1\lambda_2 + \lambda_2\lambda_3 + \lambda_1\lambda_3 \, , \\
    i_3^{\textbf{U}} &= \quad\;\; \textrm{Det}\,\mathbf{U} \qquad\qquad\quad\,\, = \lambda_1\lambda_2\lambda_3 \, , \nonumber
\end{align}
where the principal stretches $\lambda_k$ are the eigenvalues of $\mathbf{U}$.  Since $\mathbf{U}$ is positive-definite, its eigenvalues and principal invariants are all positive. 
The stretch tensors $\mathbf{U}$ and $\mathbf{V}$ have the same eigenvalues and invariants.
The Cauchy-Green strain tensors $\mathbf{C}$ and $\mathbf{B}$ also share eigenvalues, which are the squares $\lambda_k^2$ of those of $\mathbf{U}$, and invariants $i_k^{\textbf{C}}$, constructed in an analogous manner to those of $\mathbf{U}$.
Just like $\mathbf{C}$ itself, all three $i_k^{\textbf{C}}$ can be written in terms of derivatives of position. 
In our notation, Ting's expression is
\begin{equation}
	\mathbf{U} = \left( i_1^{\textbf{U}}i_2^{\textbf{U}}-i_3^{\textbf{U}} \right)^{-1}
     \left( i_1^{\textbf{U}}i_3^{\textbf{U}}G_{IJ}+\left[ \left( i_1^{\textbf{U}} \right)^2
     -i_2^{\textbf{U}} \right]g_{ij}-g_{ik}G^{KL}g_{lj} \right) \mathbf{G}^I\mathbf{G}^J \, .
    \label{eq:uexp}
\end{equation}
Similarly, the left stretch $\mathbf{V}$ may be written 
\begin{eqnarray}
	\mathbf{V} = \left( i_1^{\textbf{U}}i_2^{\textbf{U}}-i_3^{\textbf{U}} \right)^{-1}
     \left( i_1^{\textbf{U}}i_3^{\textbf{U}}g^{ij}+\left[ \left( i_1^{\textbf{U}} \right)^2
     -i_2^{\textbf{U}} \right]G^{IJ}-G^{IK}g_{kl}G^{LJ} \right) \mathbf{g}_i\mathbf{g}_j  \, .
    \label{eq:vexp}
\end{eqnarray}
The relations $\mathbf{Q} = \mathbf{Q}^{-\top} = (\bg_i\bG^I\cdot\mathbf{U})^{-\top} = \bg^i\bG_I\cdot\mathbf{U}$ 
yield the rotation
\begin{equation}
	  \mathbf{Q} = \left( i_1^{\textbf{U}}i_2^{\textbf{U}}-i_3^{\textbf{U}} \right)^{-1}
     \left( i_1^{\textbf{U}}i_3^{\textbf{U}}G_{KJ}g^{ki} +\left[ \left( i_1^{\textbf{U}} \right)^2
     -i_2^{\textbf{U}} \right] \delta^i_j -G^{IK}g_{kj} \right) \mathbf{g}_i\mathbf{G}^J \;.      
     \label{eq:qexp}
\end{equation}
The issue now becomes how to write the invariants $i_k^{\textbf{U}}$ of $\mathbf{U}$.  The third invariant is simple, as $i_3^{\textbf{U}} = \sqrt{i_3^{\textbf{C}}} = \sqrt{g/G}\,$. 
However, to obtain the other two invariants requires finding the eigenvalues of $\mathbf{C}$, or solving a quartic equation for $i_1^{\textbf{U}}$ obtained from the trace of \eqref{eq:uexp} \cite{hoger1984determination, stickforth1987square}.

In the following section, we show that neglect of terms higher order than quadratic in Biot strain allows for the computation of relevant strain invariants from a quadratic equation in which the single relevant root is easily identifiable.
This leads to explicit approximate representations of stretch and rotation in terms of derivatives of position.

\subsection{Quadratic-Biot theory: explicit approximate representation of stretch and rotation in terms of derivatives of position}
\label{sec:expquad}

The central quantity in our derivations will be the Biot strain
\begin{equation}\label{Biotdefinition}
    \bEB = \mathbf{U} - \mathbf{I} \, ,
\end{equation}
whose eigenvalues are $\Delta_k = \lambda_k-1$ and whose principal invariants are related to those of the right stretch by
\begin{align}  \label{eq:inve}
  i_1^{\textbf{E}_\text{B}} &= \Delta_1 + \Delta_2 + \Delta_3 \qquad\quad\;\;\;=\qquad\qquad\;\;\; i_1^{\textbf{U}} - 3 \, ,   \nonumber
 \\
  i_2^{\textbf{E}_\text{B}} &= \Delta_1\Delta_2+\Delta_1\Delta_3+\Delta_2\Delta_3
     = \qquad i_2^{\textbf{U}} - 2i_1^{\textbf{U}} + 3 \, ,  \\
  i_3^{\textbf{E}_\text{B}}  &= \Delta_1\Delta_2\Delta_3 \qquad\qquad\qquad\;= i_3^{\textbf{U}}-i_2^{\textbf{U}}+i_1^{\textbf{U}} - 1\, .  \nonumber
\end{align}
The simplification we seek will come from neglect of the cubic invariant. 
Note that 
\begin{equation}
    i_3^{\textbf{U}} = i_3^{\textbf{E}_\text{B}}+i_2^{\textbf{E}_\text{B}}+i_1^{\textbf{E}_\text{B}}+1 \, ,
    \label{eq:i3u}
\end{equation}
and, as $\bEB^2 = \mathbf{C} - 2\mathbf{U} + \mathbf{I}$, 
\begin{equation}
    i_2^{\textbf{E}_\text{B}} = \frac{1}{2}\Big[(\textrm{Tr}\,\bEB)^2-\textrm{Tr}\,(\bEB^2)\Big]
    = \frac{1}{2}(i_1^{\textbf{E}_\text{B}})^2 - \frac{1}{2}\,i_1^{\textbf{C}}+i_1^{\textbf{E}_\text{B}}+\frac{3}{2} \, .
\end{equation}
These expressions may be combined into a quadratic equation for the first invariant of Biot strain,
\begin{equation}
    (i_1^{\textbf{E}_\text{B}})^2 + 4i_1^{\textbf{E}_\text{B}} = i_1^{\textbf{C}}+2i_3^{\textbf{U}}-2i_3^{\textbf{E}_\text{B}}-5\;.
    \label{eq:i1e}
\end{equation}
The only term on the right hand side of \eqref{eq:i1e} that cannot be expressed in terms of derivatives of position is $i_3^{\textbf{E}_\text{B}} = \textrm{Det} (\mathbf{U}-\mathbf{I})$, but this term is $O(\Delta^3)$ (and identically zero in the case of plane strain).
Thus,
\begin{equation}
    (i_1^{\textbf{E}_\text{B}})^2 + 4i_1^{\textbf{E}_\text{B}} =  g_{ij}G^{IJ} + 2\sqrt{g/G} - 5 + O(i_3^{\textbf{E}_\text{B}}) \, ,\label{eq:i1e2}
\end{equation}
and the only relevant root is that which connects with the solution for vanishing deformation, for which $i_1^{\textbf{E}_\text{B}}$ and the right hand side of \eqref{eq:i1e2} are both zero, namely
\begin{equation}
    i_1^{\textbf{E}_\text{B}} = -2 + \sqrt{g_{ij}G^{IJ} + 2\sqrt{g/G} - 1} + O(i_3^{\textbf{E}_\text{B}}) \, .
    \label{eq:i1e-final}
\end{equation}
This result allows us to approximately express the three invariants appearing in expressions for the stretches (\ref{eq:uexp}-\ref{eq:vexp}) and rotation \eqref{eq:qexp} purely in terms of derivatives of position,
\begin{align}\label{eq:invu}
    i_1^{\textbf{U}} &= 1 + \sqrt{g_{ij}G^{IJ} + 2\sqrt{g/G} - 1} + O(i_3^{\textbf{E}_\text{B}}) \, , \nonumber \\
    i_2^{\textbf{U}} &= \sqrt{g_{ij}G^{IJ} + 2\sqrt{g/G} - 1} + \sqrt{g/G} + O(i_3^{\textbf{E}_\text{B}}) \, , \\
    i_3^{\textbf{U}} &=  \sqrt{g/G} \, . \nonumber 
\end{align}
Note that neglect of $O(\Delta^3)$ terms makes these three quantities linearly dependent, as may have been gleaned from \eqref{eq:inve}.  
The two quantities $g_{ij}G^{IJ}$ and $g/G$ appearing in \eqref{eq:invu} are the first and third invariants of $\textbf{C}$. 
Using \eqref{eq:inve} and \eqref{eq:invu} and collecting results,
\begin{align}\label{eq:inve2}
    i_1^{\textbf{E}_\text{B}} &= -2 + \sqrt{g_{ij}G^{IJ} + 2\sqrt{g/G} - 1} + O(i_3^{\textbf{E}_\text{B}}) \, , \nonumber \\
    i_2^{\textbf{E}_\text{B}} &= 1 - \sqrt{g_{ij}G^{IJ} + 2\sqrt{g/G} - 1} + \sqrt{g/G} + O(i_3^{\textbf{E}_\text{B}}) \, , \\
    i_3^{\textbf{E}_\text{B}} &=  O(\Delta_1\Delta_2\Delta_3) \, . \nonumber 
\end{align}

\section{Conservation laws and constitutive relations}
\label{sec:mec}

\subsection{Stresses}

The force per unit area in the reference and present configurations is respectively given by the tractions $\mathbf{T}$ and $\mathbf{t}$, which are related to the first Piola-Kirchhoff and Cauchy stresses $\mathbf{P}$ and $\boldsymbol\sigma$,
\begin{equation}
    \mathbf{t}\,da = \mathbf{n}\cdot\boldsymbol\sigma\,da = \boldsymbol\sigma^\top\cdot\mathbf{n}\,da
    = \sqrt{g/G}\,\boldsymbol\sigma^\top\cdot\mathbf{g}^i\mathbf{G}_I\cdot\mathbf{N}\,dA
    = \mathbf{P}\cdot\mathbf{N}\,dA = \mathbf{N}\cdot\mathbf{P}^\top dA =  \mathbf{T} \,dA\, ,
\end{equation}
where $\mathbf{N}$ and $\mathbf{n}$ are the referential and present unit normals.  The stresses are related by
$\mathbf{P} = \sqrt{g/G}\, \boldsymbol\sigma^\top\cdot\mathbf{g}^i\mathbf{G}_I$.
Because $\boldsymbol\sigma$ is often assumed symmetric, there is some inconsistency in the literature as to the definition of $\mathbf{P}$.  
For instance, both Lurie \cite{lurie1968theory} and Atluri \cite{atluri1984alternate} define $\mathbf{P}$ as the transpose of our definition.
While $\mathbf{P}$ is generally not symmetric, $\mathbf{P}\cdot\mathbf{G}^I\mathbf{g}_i$ is symmetric whenever $\boldsymbol\sigma$ is symmetric.  In Appendix~\ref{ap:pft}, we further show that $\mathbf{P}^\top\cdot\mathbf{Q}$ is always symmetric for an isotropic material with symmetric $\boldsymbol\sigma$.

\subsection{Variational framework}
\label{sec:variational}

We adopt, with slight modifications, the variational principle of Atluri and Murakawa \cite{atluri1977hybrid, atluri1984alternate, atluri1995rotations} 
to write a referential stored energy (density) $\mathcal{W}$ in terms of Biot strain, with position derivatives, rotation, and right stretch linked by a multiplier so that auxiliary fields may be varied independently.  The energy is
\begin{equation}
    \mathcal{E} (\mathbf{x},\mathbf{Q},\bEB ;  \mathbf{P})
    = \int_{\mathcal{B}} \left[ \mathcal{W} (\bEB)
    + \mathbf{P}: \left( \mathbf{g}_i\mathbf{G}^I - \mathbf{Q}\cdot\mathbf{U} \right)
    \right] dV \, .
    \label{eq:energy}
\end{equation}
Our choice of symbol anticipates the identification of the multiplier $\mathbf{P}$ with the first Piola-Kirchhoff stress.  
Three expressions are obtained from stationarity of \eqref{eq:energy} under variation of $\bx \in \mathbb{E}^3$, $\mathbf{Q} \in$ SO(3), $\mathbf{U} \in$ Sym$^+$, noting that $\delta\mathbf{U} = \delta\bEB$.
Details of the variation, and the derivation of the equations and single boundary condition on stress, are reserved for Appendix~\ref{ap:var}.
We obtain the balances of linear and angular momentum (\ref{eq:blm}-\ref{eq:bam}) and the constitutive relation \eqref{eq:biot} for the Biot stress $\boldsymbol\Sigma_{\text{Biot}}$, 
\begin{align}
	\bar{\nabla}\cdot\mathbf{P}^\top &= \mathbf{0} \, ,   \label{eq:blm} \\
	\mathbf{g}_i\mathbf{G}^I\cdot\mathbf{P}^\top &= \mathbf{P}\cdot\mathbf{G}^J\mathbf{g}_j\, ,   \label{eq:bam} \\
	\boldsymbol\Sigma_{\text{Biot}} \equiv \frac{\partial\mathcal{W}}{\partial\bEB}
     &=  \frac{1}{2}\left( \mathbf{P}^\top\cdot\mathbf{Q}+\mathbf{Q}^\top\cdot\mathbf{P} \right) \, ,   \label{eq:biot}
\end{align}
to be used alongside the compatibility constraint \eqref{eq:polar} and the restriction on rotation \eqref{eq:orthogonal}.  
An equivalent set of equations was presented by Wi{\'s}niewski \cite{wisniewski1998shell}.
One possible set of component forms is
\begin{align}
	\bar{\nabla}_J P^{iJ} = d_J P^{iJ} +\Gamma^i_{kj}P^{kJ}+\bar{\Gamma}^J_{JL}P^{iL} &= 0\, ,   \label{eq:blmc} \\
	P^{iJ} &= P^{jI} \, , \label{eq:bamc} \\
	\left( \boldsymbol\Sigma_{\text{Biot}} \right)^{IJ} &= \frac{1}{2}\left( P^{kI}Q\indices{_k^J} + Q\indices{_k^I}P^{kJ} \right)\, ,   \label{eq:biotc}
\end{align}
to accompany \eqref{eq:polar3} and \eqref{eq:orthogonal2}, where of course $Q\indices{_i^J} = g_{ik}Q\indices{^k_L}G^{LJ}$ and so on.

We have defined the Biot stress  $\boldsymbol\Sigma_{\text{Biot}}$ in a natural way, as the derivative of the stored energy $\mathcal{W}$ with respect to the Biot strain $\bEB$, rather than the more traditional \cite{ogden1997nonlinear} derivative with respect to the right stretch $\mathbf{U}$.
The Biot stress is the symmetric part of $\mathbf{P}^\top\cdot\mathbf{Q}$. The unsymmetrized quantity has a particular interpretation \cite{ogden1997nonlinear}, which can be seen from the relation
$\mathbf{N}\cdot \mathbf{P}^\top\cdot\mathbf{Q} = \mathbf{T}\cdot\mathbf{Q}$, indicating that its associated local load is a rotation of the referential traction $\mathbf{T}$.
 In Appendix~\ref{ap:bell} we discuss the Bell stress and strain, which would appear in a complementary formulation based on the left stretch $\mathbf{V}$. 

When $\mathcal{W}$ is specified, the constitutive relation \eqref{eq:biot} and, if necessary, the
angular momentum balance \eqref{eq:bam}, will provide the first Piola-Kirchhoff stress $\mathbf{P}$ in terms of $\mathbf{Q}$ and $\mathbf{U}$.  
Then the unknowns in the remaining equations are the position $\bx$, rotation $\mathbf{Q}$, and stretch $\mathbf{U}$.
In Section \ref{sec:forms}, different formulations will be presented in terms of one or two out of three of these quantities.

\section{Isotropic quadratic-Biot theory}
\label{sec:energy}

In this section we consider an elastic energy quadratic in the isotropic invariants of right stretch or, equivalently, Biot strain, 
constitutive relations for this specific energy, and forms of the field equations in terms of positions alone or combined with either rotations or stretches.

\subsection{Energy}
\label{sec:quadenergy}

A quadratic-Biot energy may be identified with that of John's two-dimensional ``harmonic'' materials \cite{john1960plane} and Lurie's ``semilinear'' materials \cite{lurie1968theory}.
This energy was considered by Neff and co-workers in \cite{neff2008symmetric}.  It also appears in Ozenda and Virga \cite{OzendaVirga21}, who however lose its desirable qualities by subsequently expanding the right stretch in powers of the Green-Lagrange strain.
Carroll \cite{carroll1988finite} introduced more general classes of energies defined by functions of stretch invariants. 
Another setting in which stretch-based energies have been explored is that of theories with independent rotational degrees of freedom, introduced either for computational convenience or to describe micropolar continua \cite{deveubeke1972new, bufler1983, bufler1985biot, makowski1990buckling,simo1992formulations,sansour1992exact, bufler1995drilling,sansour1996drilling, neff2008symmetric, Chao10}.

Consider small Biot strains around a stress-free reference configuration at which the three isotropic invariants 
$i_1^{\textbf{E}_\text{B}} = i_2^{\textbf{E}_\text{B}} = i_3^{\textbf{E}_\text{B}} = 0$.
As the third invariant is of cubic order, a general isotropic quadratic energy is of the form
\begin{align}
    \mathcal{W}\left( i_1^{\textbf{E}_\text{B}},i_2^{\textbf{E}_\text{B}} \right) &= c_1 \left( i_1^{\textbf{E}_\text{B}} \right)^2 + c_2\,i_2^{\textbf{E}_\text{B}} \, , \label{eq:wquad} \\
    &= c_1\left( -2 + \sqrt{g_{ij}G^{IJ} + 2\sqrt{g/G} - 1}\, \right)^2 + c_2 \left( 1 - \sqrt{g_{ij}G^{IJ} + 2\sqrt{g/G} - 1} + \sqrt{g/G}\, \right) + O(i_3^{\textbf{E}_\text{B}}) \, , \label{eq:wquadapprox} \\
    &=\left( c_1+\frac{c_2}{2} \right)
    \left( \textrm{Tr}\,\bEB \right)^2
    -\frac{c_2}{2}\textrm{Tr}\left(\bEB^2 \right) \, , 
\end{align}
where $c_1 \ge -c_2/2$ and $c_2 \le 0$ are constant material parameters.
In terms of an elastic tensor, with $\mathcal{W} = \frac{1}{2} \bEB:\mathbf{A}:\bEB$, we have 
$\frac{1}{2}A^{IJKL} = \left(c_1+\frac{c_2}{2}\right)G^{IJ}G^{KL} -\frac{c_2}{4}\left(G^{IK}G^{JL} + G^{IL}G^{JK}\right)$.
The approximate form \eqref{eq:wquadapprox} of the energy suggests a definition in terms of the alternate invariants 
$i_1^{\textbf{E}_\text{B}} = \sqrt{g_{ij}G^{IJ} + 2\sqrt{g/G} - 1}\,-2$ and $i_2^{\textbf{E}_\text{B}}+i_1^{\textbf{E}_\text{B}}=  \sqrt{g/G}\,-1$.
An incompressible ($\sqrt{g/G}=1$) form of the quadratic-Biot energy would depend only on the first of these, in the simplified form $\sqrt{g_{ij}G^{IJ}+1}\,-2$. 
It is illustrative to compare this with the incompressible neo-Hookean energy 
\begin{equation}
    \mathcal{W}_{\text{NH}} = {c}_{\text{NH}}\left( -3 + g_{ij}G^{IJ} \right) \, ,
\end{equation}
in which only some of the possible quadratic-stretch terms appear.
Incompressibility needs to be implemented with an additional constraint (pressure) term, but the energetic parts of quadratic-Biot  with $c_2=-4c_1$ and neo-Hookean would agree to $O(i_3^{\textbf{E}_\text{B}})$.

\subsection{Constitutive relations}

For an isotropic material with symmetric Cauchy stress $\boldsymbol\sigma$, the terms in the Biot stress are identically symmetric, so that $\boldsymbol\Sigma_{\text{Biot}} =  \mathbf{P}^\top\cdot\mathbf{Q} = \mathbf{Q}^\top\cdot\mathbf{P}$.
The first Piola-Kirchhoff stress $\mathbf{P}$  
can now be written as a function of $\mathbf{Q}$, $\mathbf{U}$, and derivatives of $\mathcal{W}$ with respect to the invariants of $\mathbf{U}$ \cite{wheeler1990derivatives, wisniewski1996note,steigmann2002invariants}. The derivation, as carefully detailed by Wheeler \cite{wheeler1990derivatives}, leads to the expression
\begin{equation}
    \mathbf{P} = \left(\frac{\partial\mathcal{W}}{\partial i_1^{\textbf{U}}}
        +i_1^{\textbf{U}}\frac{\partial\mathcal{W}}{\partial i_2^{\textbf{U}}} \right)\mathbf{Q}
    - \frac{\partial\mathcal{W}}{\partial i_2^{\textbf{U}}}\mathbf{Q}\cdot\mathbf{U}
    +\frac{\partial\mathcal{W}}{\partial i_3^{\textbf{U}}} i_3^{\textbf{U}}
    (\mathbf{Q}\cdot\mathbf{U})^{-\top} \, .
        \label{eq:pkcr}
\end{equation}
Using the Cayley-Hamilton theorem \cite{steigmann2002invariants}, $\mathbf{U}^3-i_1^{\textbf{U}}\mathbf{U}^2+i_2^{\textbf{U}}\mathbf{U}-i_3^{\textbf{U}}\mathbf{I} = \bm{0}$, which allows replacement of $\mathbf{U}^{-1}$ thus: 
\begin{equation}
    \mathbf{P} = \bigg(\frac{\partial\mathcal{W}}{\partial i_1^{\textbf{U}}}
        +i_1^{\textbf{U}}\frac{\partial\mathcal{W}}{\partial i_2^{\textbf{U}}}
        +i_2^{\textbf{U}}\frac{\partial\mathcal{W}}{\partial i_3^{\textbf{U}}}\bigg)\mathbf{Q}
    - \bigg(\frac{\partial\mathcal{W}}{\partial i_2^{\textbf{U}}}+i_1^{\textbf{U}}
        \frac{\partial\mathcal{W}}{\partial i_3^{\textbf{U}}} \bigg)\mathbf{Q}\cdot\mathbf{U}
    +\frac{\partial\mathcal{W}}{\partial i_3^{\textbf{U}}} \mathbf{Q}\cdot\mathbf{U}^2 \;.
    \label{eq:1pkiso}
\end{equation}
The relationships between invariants \eqref{eq:inve}, those between stresses, and the definition of the Biot strain, convert this expression into one for the Biot stress in terms of the Biot strain and its invariants,
\begin{eqnarray}
  &&\boldsymbol\Sigma_{\text{Biot}} = \bigg(\frac{\partial\mathcal{W}}{\partial i_1^{\textbf{E}_\text{B}}}
     +i_1^{\textbf{E}_\text{B}}\frac{\partial\mathcal{W}}{\partial i_2^{\textbf{E}_\text{B}}}
     +i_2^{\textbf{E}_\text{B}}\frac{\partial\mathcal{W}}{\partial i_3^{\textbf{E}_\text{B}}}\bigg)\mathbf{I}
     -\bigg(\frac{\partial\mathcal{W}}{\partial i_2^{\textbf{E}_\text{B}}}
     +i_1^{\textbf{E}_\text{B}}\frac{\partial\mathcal{W}}{\partial i_3^{\textbf{E}_\text{B}}} \bigg)\bEB
     +\frac{\partial\mathcal{W}}{\partial i_3^{\textbf{E}_\text{B}}}\bEB^2 \;.
\end{eqnarray}

Specifically for the quadratic-Biot energy \eqref{eq:wquad}, 
\begin{align}\label{quadraticBiotstress}
	\boldsymbol\Sigma_{\text{Biot}} &= (2c_1+c_2)i_1^{\textbf{E}_\text{B}}\,\mathbf{I}-c_2\, \bEB \, , \\
	&= (2c_1+c_2)\left( -2 + \sqrt{g_{ij}G^{IJ} + 2\sqrt{g/G} - 1}\, \right) \mathbf{I}-c_2\, \bEB + O(i_3^{\textbf{E}_\text{B}}) \, ,  \nonumber 
\end{align}
and $\mathbf{P} = \mathbf{Q}\cdot\boldsymbol\Sigma_{\text{Biot}}$.

\subsection{Formulations in terms of rotations or stretches}
\label{sec:forms}

Either the rotation or stretch field may be eliminated to form a set of equations for two fields, one of which must be obtained by solving a quadratic equation, or both rotation and stretch can be eliminated to obtain an approximate description in terms of position derivatives alone.
Formulations involving independent rotations have been developed in \cite{simo1992formulations, wisniewski1998shell, merlini1997variational}. 
While these were motivated by computational concerns, the equations presented below are relatively simple from an analytical point of view.

A description in terms of position derivatives and rotations is the linear momentum equation \eqref{eq:blm} with
\begin{align}  
\label{eq:qblm}
	\mathbf{P} &= \left[ \left( 2c_1+c_2 \right)  i_1^{\textbf{E}_\text{B}} + c_2 \right] \mathbf{Q} - c_2\,\mathbf{g}_i\mathbf{G}^I \, , \\
	P^{iJ} &=  \left[ \left( 2c_1+c_2 \right)  i_1^{\textbf{E}_\text{B}} + c_2 \right]Q\indices{^i_K}G^{KJ}-c_2G^{IJ} \;,
\end{align}
with  $i_1^{\textbf{E}_\text{B}} = Q\indices{_k^K} -3 = g_{ki}Q\indices{^i_J}G^{KJ}-3 = -2 + \sqrt{g_{ij}G^{IJ} + 2\sqrt{g/G} - 1} + O(i_3^{\textbf{E}_\text{B}})$.  The rotations may be obtained either exactly by solving the constraint \eqref{eq:orthogonal} or \eqref{eq:orthogonal2}, or approximately by using the explicit form from \eqref{eq:qexp} and \eqref{eq:invu}. 

A description in terms of position derivatives and stretches is the linear momentum equation \eqref{eq:blm} with
\begin{align}  
\label{eq:ublm}
	\mathbf{P} &= \left[ \left( 2c_1+c_2 \right)  i_1^{\textbf{E}_\text{B}} + c_2 \right] \bg^i\bG_I\cdot\mathbf{U} - c_2\,\mathbf{g}_i\mathbf{G}^I \, , \\
	P^{iJ} &=  \left[ \left( 2c_1+c_2 \right)  i_1^{\textbf{E}_\text{B}} + c_2 \right]g^{ik}U_{KL}G^{LJ}-c_2G^{IJ} \;,
\end{align}
with  $i_1^{\textbf{E}_\text{B}} = U\indices{^I_I} -3 = G^{IJ}U_{JI}-3 = -2 + \sqrt{g_{ij}G^{IJ} + 2\sqrt{g/G} - 1} + O(i_3^{\textbf{E}_\text{B}})$. We have used $\mathbf{Q}= \mathbf{Q}^{-\top}$ to rewrite $\mathbf{P}$ in terms of $\mathbf{U}$. 
The stretches may be obtained either exactly by solving \eqref{eq:Usquared}, or approximately by using the explicit form from \eqref{eq:uexp} and \eqref{eq:invu}.

\section{Towards anisotropic theories}
\label{sec:ani}

Many soft material structures of the type our theory is intended to address are anisotropic.
In this section, we briefly indicate how to construct energies and balance equations for such materials, through the example of a transversely isotropic elastic solid such as a fiber-reinforced material.

Let the anisotropic material have a distinguished direction, in the reference configuration, identified with the unit vector $\mathbf{D}$. 
Following the pattern for constructing a transversely isotropic energy using this quantity and a strain tensor  \cite{spencer1984constitutive,murphy2013transversely}, we obtain to quadratic order in stretch
\begin{align}
  \mathcal{W}(\bEB,\mathbf{D}) = \left(c_1+\frac{c_2}{2}\right)\left( \textrm{Tr}\,\bEB \right)^2
     -\frac{c_2}{2}\textrm{Tr}\left(\bEB^2 \right)
     + c_3\left( \textrm{Tr}\,\bEB \right)(\bEB:\mathbf{D}\mathbf{D})
     + c_4(\bEB:\mathbf{D}\mathbf{D})^2
     + c_5\bEB^2:\mathbf{D}\mathbf{D} \, ,
\end{align}
where the $c_i$ are constant material parameters.  
In terms of an elastic tensor, with $\mathcal{W} = \frac{1}{2} \bEB:\mathbf{A}:\bEB$, we have 
\newline $\frac{1}{2}A^{IJKL} = \left(c_1+\frac{c_2}{2}\right)G^{IJ}G^{KL}
     -\frac{c_2}{4}\left(G^{IK}G^{JL}+G^{IL}G^{JK}\right)
     +c_3G^{IJ}D^KD^L +c_4D^ID^JD^KD^L + \newline \frac{c_5}{4}\left(G^{IK}D^JD^L+G^{IL}D^JD^K + G^{JK}D^ID^L+G^{JL}D^ID^K\right)$.
The Biot stress is still computed as the derivative of $\mathcal{W}$ with respect to $\bEB$,
\begin{align}
	\boldsymbol\Sigma_{\text{Biot}} = (2c_1+c_2)i_1^{\textbf{E}_\text{B}}\mathbf{I}-c_2\bEB
  +c_3\left( \mathbf{I}\,\bEB:\mathbf{D}\mathbf{D}+i_1^{\textbf{E}_\text{B}}\mathbf{D}\mathbf{D} \right)
     +2c_4\bEB:\mathbf{D}\mathbf{D}\,\mathbf{D}\mathbf{D}
       +c_5\left( \mathbf{D}\mathbf{D}\cdot\bEB+\bEB\cdot\mathbf{D}\mathbf{D} \right) \, ,
\end{align}
but $\mathbf{P}^\top\cdot\mathbf{Q}$ is no longer symmetric.  The first Piola-Kirchhoff stress can be written \cite{wisniewski1998shell} in terms of the symmetric Biot stress plus an additional anisotropic contribution $\boldsymbol\Sigma_\text{an}$,
\begin{equation}
    \mathbf{P} = \mathbf{Q}\cdot \left( \boldsymbol\Sigma_{\text{Biot}}+\boldsymbol\Sigma_\text{an} \right)\, ,
\end{equation}
whose determination will require use of the angular momentum balance \eqref{eq:bam} alongside the linear momentum balance \eqref{eq:blm} and constitutive equation \eqref{eq:biot}.

\section{Conclusions}\label{sec:conc}

This paper has established a basis for small-strain nonlinear-elastic theories with energies quadratic in stretch, reflecting a systematic expansion in Biot strains.  Results on the kinematics of stretch and rotation, and variational principles for elasticity with auxiliary fields, have been further developed and combined within this framework.
Neglect of higher order strains results in simple algebraic expressions for isotropic invariants that bypass the need for complex operations involving quartic equations or tensor square roots.
Stresses, balance laws, and constitutive relations are expressed in terms of derivatives of position, with optional simultaneous consideration of a stretch or rotation field.
The ideas are developed in the context of an isotropic material, with a brief sketch of anisotropic extensions.

\section*{Acknowledgments}

This work was supported by U.S. National Science Foundation grant CMMI-2001262.  Helpful conversations with O. Oshri, D. J. Steigmann, E. G. Virga, and A. Yavari are acknowledged.

\appendix

\section{Other decompositions of the deformation gradient}
\label{ap:dg}

The multiplicative polar decomposition of the deformation gradient \eqref{eq:polar} is a cornerstone of continuum mechanics, possessing several desirable features, including unique symmetric positive-definite right and left stretches.
However, it is not the only option.
In micropolar continuum theories \cite{steinmann1997unifying, merlini1997variational, pietraszkiewicz2009natural} the multiplicative decomposition involves an independent micropolar rotation and a non-symmetric stretch.
Chen \cite{Chen86} 
proposed an additive decomposition into a symmetric strain tensor and an orthogonal rotation tensor,
\begin{equation}
    \mathbf{g}_i\mathbf{G}^I = \mathbf{S} + \mathbf{R} \, . \label{SRdecomp}
\end{equation}
The strain $\mathbf{S}$ is unique.
By contrast, there are two choices in the polar decomposition, corresponding to the order of application of strain and rotation, with corresponding natural notions of referential (Biot) and present (Bell) strains.
The tensors in the decomposition \eqref{SRdecomp} can be computed directly from derivatives of position  \cite{Chen86},
\begin{align}
	\mathbf{S} &= -\mathbf{I} + \frac{1}{2}\left(\mathbf{g}_i\mathbf{G}^I+\mathbf{G}^I\mathbf{g}_i\right)
     -\frac{1}{1+\textrm{cos}\theta}\mathbf{W}\cdot\mathbf{W} \, , \\
	\mathbf{R} &= \mathbf{I}+\frac{1}{2}\mathbf{W}
     +\frac{1}{1+\textrm{cos}\theta}\mathbf{W}\cdot\mathbf{W} \, ,  \\
	\mathbf{W} &= \frac{1}{2}\left(\mathbf{g}_i\mathbf{G}^I-\mathbf{G}^I\mathbf{g}_i\right) \, , \\
	\textrm{cos}\theta  &= \left(1-\mathbf{W}:\mathbf{W}\right)^{1/2} \, ,
\end{align}
However, in contrast to the stretches 
 from the polar decomposition, the strain $\mathbf{S}$ is not a pure measure of material distortion but is still corrupted by irrelevant rotational information, and the functional form of an objective elastic energy must include some combination of both strain $\mathbf{S}$ and rotation $\mathbf{R}$.
 This approach has, to our knowledge, not been explored. 
In micropolar theories \cite{steinmann1997unifying, merlini1997variational, pietraszkiewicz2009natural}, a quantity of the type $\mathbf{S} = \mathbf{g}_i\mathbf{G}^I-\mathbf{R}$ is one possible definition for the linear strain, to be accompanied by an angular strain.

\section{Symmetry of $\mathbf{P}^\top\cdot\mathbf{Q}$ for isotropic materials}
\label{ap:pft}

The following line of reasoning may be found in Lurie \cite{lurie1968theory} and Ogden \cite{ogden1997nonlinear}.
For isotropic materials with symmetric Cauchy stress $\boldsymbol\sigma$, there exists a representation $\boldsymbol\sigma = c_0\mathbf{I} + c_1\mathbf{V}+c_2\mathbf{V}^2$, 
where the $c_i$ are functions of the invariants of $\mathbf{V}$. 
This means that $\boldsymbol\sigma$ is coaxial to (shares eigenvectors with) $\mathbf{V}$ and $\mathbf{B}$ and, consequently, $\mathbf{Q}^\top\cdot\boldsymbol\sigma\cdot\mathbf{Q}$ is coaxial to $\mathbf{U}$ and $\mathbf{C}$. 
Noting further that $\mathbf{U}$ is coaxial with its inverse, and that 
$\mathbf{P}^\top\cdot\mathbf{Q} = \sqrt{g/G}\,\mathbf{U}^{-1}\cdot\mathbf{Q}^\top\cdot\boldsymbol\sigma\cdot\mathbf{Q}$ and $\mathbf{Q}^\top\cdot\mathbf{P} = \sqrt{g/G}\,\mathbf{Q}^\top\cdot\boldsymbol\sigma\cdot\mathbf{Q}\cdot\mathbf{U}^{-1}$, 
the coaxiality of $\mathbf{Q}^\top\cdot\boldsymbol\sigma\cdot\mathbf{Q}$ and $\mathbf{U}^{-1}$ implies the symmetry of $\mathbf{P}^\top\cdot\mathbf{Q}$, which can thus be identified with the Biot stress \emph{via} \eqref{eq:biot}.

\section{Variation of the energy}
\label{ap:var}

Here we detail the first variation of the energy \eqref{eq:energy} under independent shifts in $\bx \in \mathbb{E}^3$, $\mathbf{Q} \in$ SO(3), $\mathbf{U} \in$ Sym$^+$.  The approach follows Atluri and Murakawa \cite{atluri1977hybrid}.

Noting that $\delta\mathbf{U} = \delta\bEB$ and using $\mathbf{P}: \delta\mathbf{Q} \cdot\mathbf{U} = \delta\mathbf{Q} : \mathbf{P}\cdot\mathbf{U}$
and $\mathbf{P}:\mathbf{Q}\cdot\delta\bEB = \mathbf{Q}^\top\cdot\mathbf{P}:\delta\bEB$, we obtain
\begin{align}
  \delta\mathcal{E}
  = \int_{\mathcal{B}} \left[
  -\bar{\nabla}\cdot\mathbf{P}^\top\cdot\delta\mathbf{x}
  - \delta\mathbf{Q} : \mathbf{P}\cdot\mathbf{U} 
  +\left( \frac{\partial\mathcal{W}}{\partial\bEB}
      -  \mathbf{Q}^\top \cdot \mathbf{P} \right)
      :\delta\bEB 
     \right] dV
     +\int_{\partial\mathcal{B}}\mathbf{N}\cdot\mathbf{P}^\top\cdot\delta\mathbf{x}\,dA \, .
     \label{eq:1var}
\end{align}

The terms conjugate to $\delta\mathbf{x}$ directly provide the linear momentum balance \eqref{eq:blm} and a boundary condition $\mathbf{N}\cdot\mathbf{P}^\top = \bm{0}$.

Using $\delta\mathbf{Q}\cdot\mathbf{Q}^\top = -\mathbf{Q}\cdot\delta\mathbf{Q}^\top$, we rewrite the quantity involving $\delta\mathbf{Q}$ as $-\delta\mathbf{Q} : \mathbf{P}\cdot\mathbf{U} = \left(\mathbf{Q} \cdot \delta\mathbf{Q}^\top \cdot \mathbf{Q} \right) : \mathbf{P}\cdot\mathbf{U} =  \delta\mathbf{Q}^\top \cdot \mathbf{Q} : \left( \mathbf{Q}^\top \cdot \mathbf{P}\cdot\mathbf{U} \right)$.  Because
$\delta\mathbf{Q}^\top\cdot\mathbf{Q}$ is antisymmetric, only the antisymmetric part of the conjugate quantity need vanish,
\begin{align}
	\mathbf{U} \cdot \mathbf{P}^\top \cdot \mathbf{Q} = \mathbf{Q}^\top \cdot \mathbf{P}\cdot\mathbf{U} \, .
\end{align}
Left dotting with $\mathbf{Q}$ and right dotting with $\mathbf{Q}^\top$ provides the form of the angular momentum balance shown in \eqref{eq:bam}.

Because $\delta\bEB$ is symmetric, only the symmetric part of its conjugate quantity need vanish, leading to the constitutive equation \eqref{eq:biot}.

\section{Bell strain and stress}
\label{ap:bell}

Our framework has employed the referential right stretch and associated Biot strain.  
Their counterparts are the present left stretch $\mathbf{V}$ and the Bell strain, defined as
\begin{equation}
    \mathbf{E}_{\text{Bell}} = \mathbf{V}-\mathbf{I} \, .
    \label{eq:bell}
\end{equation}
The Biot and Bell strains share eigenvalues and invariants. 
A variational principle based on the energy
\begin{equation}
    \mathcal{E} (\mathbf{x},\mathbf{Q},\mathbf{E}_{\text{Bell}} ;  \mathbf{P})
    = \int_{\mathcal{B}} \left[\mathcal{W} (\mathbf{E}_{\text{Bell}})
    + \mathbf{P}:\left(\mathbf{g}_i\mathbf{G}^I
    -\mathbf{V}\cdot\mathbf{Q}\right)
    \right]dV 
    \label{eq:energy2}
\end{equation}
would result in a constitutive relation for the symmetric Bell stress \cite{bell1986continuum,beatty1992deformations,steigmann2002invariants},
\begin{equation}
    \boldsymbol\Sigma_{\text{Bell}} \equiv
    \frac{\partial\mathcal{W}}{\partial\mathbf{E}_{\text{Bell}}}
    = \frac{1}{2}\left( \mathbf{Q}\cdot\mathbf{P}^\top+\mathbf{P}\cdot\mathbf{Q}^\top \right)\, .
\end{equation}
For isotropic materials with symmetric Cauchy stress, $\boldsymbol\Sigma_{\text{Biot}} = \mathbf{P}^\top\cdot\mathbf{Q}$ and $\boldsymbol\Sigma_{\text{Bell}} = \mathbf{Q}\cdot\mathbf{P}^\top$, so the first Piola-Kirchhoff stress $\mathbf{P}$ admits the decomposition
\begin{equation}
    \mathbf{P} = \mathbf{Q}\cdot\boldsymbol\Sigma_{\text{Biot}} = \boldsymbol\Sigma_{\text{Bell}}\cdot\mathbf{Q} \, .
\end{equation} 
Despite its resemblance to the polar decomposition of the deformation gradient, this decomposition for $\mathbf{P}$ is not unique, since $\boldsymbol\Sigma_{\text{Biot}}$ and $\boldsymbol\Sigma_{\text{Bell}}$ are generally neither positive- nor negative-definite.

\bibliographystyle{unsrt}

\begin{thebibliography}{10}

\bibitem{ogden1997nonlinear}
R.~W. Ogden.
\newblock {\em Non-Linear Elastic Deformations}.
\newblock Courier, 1997.

\bibitem{Efrati09jmps}
E.~Efrati, E.~Sharon, and R.~Kupferman.
\newblock Elastic theory of unconstrained non-{E}uclidean plates.
\newblock {\em Journal of the Mechanics and Physics of Solids}, 57:762--775,
  2009.

\bibitem{Dias11}
M.~A. Dias, J.~A. Hanna, and C.~D. Santangelo.
\newblock Programmed buckling by controlled lateral swelling in a thin elastic
  sheet.
\newblock {\em Physical Review E}, 84:036603, 2011.

\bibitem{Pezzulla15}
M.~Pezzulla, N.~Stoop, X.~Jiang, and D.~P. Holmes.
\newblock Morphing of geometric composites \emph{via} residual swelling.
\newblock {\em Soft Matter}, 11:5812--5820, 2015.

\bibitem{NguyenSelinger17}
T.-S. Nguyen and J.~V. Selinger.
\newblock Theory of liquid crystal elastomers and polymer networks.
\newblock {\em The European Physical Journal E}, 40:76, 2017.

\bibitem{vanRees17}
W.~M. van Rees, E.~Vouga, and L.~Mahadevan.
\newblock Growth patterns for shape-shifting elastic bilayers.
\newblock {\em PNAS}, 114(44):11597--11602, 2017.

\bibitem{IrschikGerstmayr09}
H.~Irschik and J.~Gerstmayr.
\newblock A continuum mechanics based derivation of {R}eissner's
  large-displacement finite-strain beam theory: the case of plane deformations
  of originally straight {B}ernoulli-{E}uler beams.
\newblock {\em Acta Mechanica}, 206:1--21, 2009.

\bibitem{OshriDiamant17}
O.~Oshri and H.~Diamant.
\newblock Strain tensor selection and the elastic theory of incompatible thin
  sheets.
\newblock {\em Physical Review E}, 95:053003, 2017.

\bibitem{Hanna19}
J.~A. Hanna.
\newblock Some observations on variational elasticity and its application to
  plates and membranes.
\newblock {\em Zeitschrift f{\"{u}}r angewandte Mathematik und Physik}, 70:76,
  2019.

\bibitem{WoodHanna19}
H.~G. Wood and J.~A. Hanna.
\newblock Contrasting bending energies from bulk elastic theories.
\newblock {\em Soft Matter}, 15:2411--2417, 2019.

\bibitem{OzendaVirga21}
O.~Ozenda and E.~G. Virga.
\newblock On the {K}irchhoff-{L}ove hypothesis (revised and vindicated).
\newblock {\em Journal of Elasticity}, 143:359--384, 2021.

\bibitem{Antman68-2}
S.~Antman.
\newblock General solutions for plane extensible elasticae having nonlinear
  stress-strain laws.
\newblock {\em Quarterly of Applied Mathematics}, 26(1):35--47, 1968.

\bibitem{Reissner72}
E.~Reissner.
\newblock On one-dimensional finite-strain beam theory: the plane problem.
\newblock {\em Zeitschrift f{\"{u}}r angewandte Mathematik und Physik},
  23:795--804, 1972.

\bibitem{WhitmanDeSilva74}
A.~B. Whitman and C.~N. DeSilva.
\newblock An exact solution in a nonlinear theory of rods.
\newblock {\em Journal of Elasticity}, 4(4):265--280, 1974.

\bibitem{IwakumaKuranishi84}
T.~Iwakuma and S.~Kuranishi.
\newblock How much contribution does the shear deformation have in a beam
  theory?
\newblock {\em Structural Engineering/Earthquake Engineering, Proceedings of
  the Japan Society of Civil Engineers}, 344:141--151, 1984.

\bibitem{Chaisomphob86}
T.~Chaisomphob, F.~Nishino, A.~Hasegawa, and A.~G.~A. Abdel-Shafy.
\newblock An elastic finite displacement analysis of plane beams with and
  without shear deformation.
\newblock {\em Structural Engineering/Earthquake Engineering, Proceedings of
  the Japan Society of Civil Engineers}, 368:169--177, 1986.

\bibitem{Magnusson01}
A.~Magnusson, M.~Ristinmaa, and C.~Ljung.
\newblock Behaviour of the extensible elastica solution.
\newblock {\em International Journal of Solids and Structures}, 38:8441--8457,
  2001.

\bibitem{Antman05}
S.~S. Antman.
\newblock {\em Nonlinear Problems of Elasticity}.
\newblock Springer, New York, second edition, 2005.

\bibitem{KnocheKierfeld11}
S.~Knoche and J.~Kierfeld.
\newblock Buckling of spherical capsules.
\newblock {\em Physical Review E}, 84:046608, 2011.

\bibitem{SeungNelson88}
H.~S. Seung and D.~R. Nelson.
\newblock Defects in flexible membranes with crystalline order.
\newblock {\em Physical Review A}, 38(2):1005--1018, 1988.

\bibitem{Deserno15}
M.~Deserno.
\newblock Fluid lipid membranes: From differential geometry to curvature
  stresses.
\newblock {\em Chemistry and Physics of Lipids}, 185:11--45, 2015.

\bibitem{lurie1968theory}
A.~I. Lur\textsf{'}e.
\newblock Theory of elasticity for a semilinear material.
\newblock {\em Journal of Applied Mathematics and Mechanics}, 32(6):1068--1085,
  1968.

\bibitem{john1960plane}
F.~John.
\newblock Plane strain problems for a perfectly elastic material of harmonic
  type.
\newblock {\em Communications on Pure and Applied Mathematics}, 13(2):239--296,
  1960.

\bibitem{atluri1977hybrid}
S.~N. Atluri and H.~Murakawa.
\newblock On hybrid finite element models in nonlinear solid mechanics.
\newblock In P.G.~Bergan et~al., editor, {\em Finite Elements in Nonlinear
  Mechanics}, pages 3--41. Tapir, 1977.

\bibitem{wisniewski1998shell}
K.~Wisniewski.
\newblock A shell theory with independent rotations for relaxed {B}iot stress
  and right stretch strain.
\newblock {\em Computational Mechanics}, 21(2):101--122, 1998.

\bibitem{merlini1997variational}
T.~Merlini.
\newblock A variational formulation for finite elasticity with independent
  rotation and {B}iot-axial fields.
\newblock {\em Computational Mechanics}, 19(3):153--168, 1997.

\bibitem{Modes11}
C.~D. Modes, K.~Bhattacharya, and M.~Warner.
\newblock Gaussian curvature from flat elastica sheets.
\newblock {\em Proceedings of the Royal Society A}, 467:1121--1140, 2011.

\bibitem{Plucinsky18}
P.~Plucinsky, B.~A. Kowalski, T.~J. White, and K.~Bhattacharya.
\newblock Patterning nonisometric origami in nematic elastomer sheets.
\newblock {\em Soft Matter}, 14:3127--3134, 2018.

\bibitem{Steigmann90}
D.~J. Steigmann.
\newblock Tension-field theory.
\newblock {\em Proceedings of the Royal Society of London A}, 429:141--173,
  1990.

\bibitem{SadikYavari17}
S.~Sadik and A.~Yavari.
\newblock On the origins of the idea of the multiplicative decomposition of the
  deformation gradient.
\newblock {\em Mathematics and Mechanics of Solids}, 22:771--772, 2017.

\bibitem{Nardinocchi13}
P.~Nardinocchi, L.~Teresi, and V.~Varano.
\newblock The elastic metric: A review of elasticity with large distortions.
\newblock {\em International Journal of Non-Linear Mechanics}, 56:34--42, 2013.

\bibitem{ericksen1957exact}
J.~L. Ericksen and C.~Truesdell.
\newblock Exact theory of stress and strain in rods and shells.
\newblock {\em Archive for Rational Mechanics and Analysis}, 1(1):295--323,
  1957.

\bibitem{malvern1969introduction}
L.~E. Malvern.
\newblock {\em Introduction to Continuum Mechanics}.
\newblock Prentice Hall, 1969.

\bibitem{stephenson1980uniqueness}
R.~A. Stephenson.
\newblock On the uniqueness of the square-root of a symmetric,
  positive-definite tensor.
\newblock {\em Journal of Elasticity}, 10(2), 1980.

\bibitem{biot1965mechanics}
M.~A. Biot.
\newblock {\em Mechanics of Incremental Deformations}.
\newblock John Wiley \& Sons, 1965.

\bibitem{hoger1984determination}
A.~Hoger and D.~E. Carlson.
\newblock Determination of the stretch and rotation in the polar decomposition
  of the deformation gradient.
\newblock {\em Quarterly of Applied Mathematics}, 42(1):113--117, 1984.

\bibitem{Sawyers86}
K.~Sawyers.
\newblock Comments on the paper \emph{Determination of the stretch and rotation
  in the polar decomposition of the deformation gradient} by {A}. {H}oger and
  {D}. {E}. {C}arlson.
\newblock {\em Quarterly of Applied Mathematics}, 44(2):309--311, 1986.

\bibitem{ting1985determination}
T.~C.~T. Ting.
\newblock Determination of {C}$^{1/2}$, {C}$^{-1/2}$ and more general isotropic
  tensor functions of {C}.
\newblock {\em Journal of Elasticity}, 15(3):319--323, 1985.

\bibitem{stickforth1987square}
J.~Stickforth.
\newblock The square root of a three-dimensional positive tensor.
\newblock {\em Acta Mechanica}, 67(1-4):233--235, 1987.

\bibitem{lu1997direct}
J.~Lu and P.~Papadopoulos.
\newblock On the direct determination of the rotation tensor from the
  deformation gradient.
\newblock {\em Mathematics and Mechanics of Solids}, 2(1):17--26, 1997.

\bibitem{guan1998determination}
D.~Guan-Suo.
\newblock Determination of the rotation tensor in the polar decomposition.
\newblock {\em Journal of Elasticity}, 50(3):197--207, 1998.

\bibitem{jog2002explicit}
C.~S. Jog.
\newblock On the explicit determination of the polar decomposition in
  $n$-dimensional vector spaces.
\newblock {\em Journal of Elasticity}, 66(2):159--169, 2002.

\bibitem{bouby2005direct}
C.~Bouby, D.~Fortun{\'e}, W.~Pietraszkiewicz, and C.~Vall{\'e}e.
\newblock Direct determination of the rotation in the polar decomposition of
  the deformation gradient by maximizing a {R}ayleigh quotient.
\newblock {\em Zeitschrift f{\"u}r Angewandte Mathematik und Mechanik},
  85(3):155--162, 2005.

\bibitem{norris2007invariants}
A.~Norris.
\newblock Invariants of {C}$^{1/2}$ in terms of the invariants of {C}.
\newblock {\em Journal of Mechanics of Materials and Structures},
  2(9):1805--1812, 2007.

\bibitem{scott2020u}
N.~H. Scott.
\newblock {U} = {C}$^{1/2}$ and its invariants in terms of {C} and its
  invariants.
\newblock {\em Journal of Elasticity}, 141(2):363--379, 2020.

\bibitem{sun2021explicit}
B.-H. Sun.
\newblock Explicit representation for the {SO(3)} rotation tensor of deformable
  bodies.
\newblock {\em Applied Mathematics Letters}, 111:106606, 2021.

\bibitem{atluri1984alternate}
S.~N. Atluri.
\newblock Alternate stress and conjugate strain measures, and mixed variational
  formulations involving rigid rotations, for computational analyses of
  finitely deformed solids, with application to plates and shells—{I}:
  {T}heory.
\newblock {\em Computers \& Structures}, 18(1):93--116, 1984.

\bibitem{atluri1995rotations}
S.~N. Atluri and A.~Cazzani.
\newblock Rotations in computational solid mechanics.
\newblock {\em Archives of Computational Methods in Engineering}, 2(1):49--138,
  1995.

\bibitem{neff2008symmetric}
P.~Neff, A.~Fischle, and I.~M{\"u}nch.
\newblock Symmetric {C}auchy stresses do not imply symmetric {B}iot strains in
  weak formulations of isotropic hyperelasticity with rotational degrees of
  freedom.
\newblock {\em Acta Mechanica}, 197(1):19--30, 2008.

\bibitem{carroll1988finite}
M.~M. Carroll.
\newblock Finite strain solutions in compressible isotropic elasticity.
\newblock {\em Journal of Elasticity}, 20(1):65--92, 1988.

\bibitem{deveubeke1972new}
B.~Fraeijs de~Veubeke.
\newblock A new variational principle for finite elastic displacements.
\newblock {\em International Journal of Engineering Science}, 10(9):745--763,
  1972.

\bibitem{bufler1983}
H.~Bufler.
\newblock On the work theorems for finite and incremental elastic deformations
  with discontinuous fields: a unified treatment of different versions.
\newblock {\em Computer Methods in Applied Mechanics and Engineering},
  36:95--124, 1983.

\bibitem{bufler1985biot}
H.~Bufler.
\newblock The {B}iot stresses in nonlinear elasticity and the associated
  generalized variational principles.
\newblock {\em Ingenieur-Archiv}, 55(6):450--462, 1985.

\bibitem{makowski1990buckling}
J.~Makowski and H.~Stumpf.
\newblock Buckling equations for elastic shells with rotational degrees of
  freedom undergoing finite strain deformation.
\newblock {\em International Journal of Solids and Structures}, 26(3):353--368,
  1990.

\bibitem{simo1992formulations}
J.~C. Simo, D.~D. Fox, and T.~J.~R. Hughes.
\newblock Formulations of finite elasticity with independent rotations.
\newblock {\em Computer Methods in Applied Mechanics and Engineering},
  95(2):277--288, 1992.

\bibitem{sansour1992exact}
C.~Sansour and H.~Bufler.
\newblock An exact finite rotation shell theory, its mixed variational
  formulation and its finite element implementation.
\newblock {\em International Journal for Numerical Methods in Engineering},
  34(1):73--115, 1992.

\bibitem{bufler1995drilling}
H.~Bufler.
\newblock On drilling degrees of freedom in nonlinear elasticity and a
  hyperelastic material description in terms of the stretch tensor. {P}art 1:
  {T}heory.
\newblock {\em Acta Mechanica}, 113(1):21--35, 1995.

\bibitem{sansour1996drilling}
C.~Sansour, H.~Bufler, and H.~M{\"u}llersch{\"o}n.
\newblock On drilling degrees of freedom in nonlinear elasticity and a
  hyperelastic material description in terms of the stretch tensor. {P}art 2:
  {A}pplication to membranes.
\newblock {\em Acta Mechanica}, 115(1):103--117, 1996.

\bibitem{Chao10}
I.~Chao, U.~Pinkall, P.~Sanan, and P.~Schr{\"{o}}der.
\newblock A simple geometric model for elastic deformations.
\newblock {\em ACM Transactions on Graphics}, 29(4):Article 38, 2010.

\bibitem{wheeler1990derivatives}
L.~T. Wheeler.
\newblock On the derivatives of the stretch and rotation with respect to the
  deformation gradient.
\newblock {\em Journal of Elasticity}, 24(1-3):129--133, 1990.

\bibitem{wisniewski1996note}
K.~Wi{\'s}niewski and E.~Turska.
\newblock A note on the hyperelastic constitutive equation for rotated {B}iot
  stress.
\newblock {\em Archives of Mechanics}, 48(5):947--953, 1996.

\bibitem{steigmann2002invariants}
D.~J. Steigmann.
\newblock Invariants of the stretch tensors and their application to finite
  elasticity theory.
\newblock {\em Mathematics and Mechanics of Solids}, 7(4):393--404, 2002.

\bibitem{spencer1984constitutive}
A.~J.~M. Spencer.
\newblock Constitutive theory for strongly anisotropic solids.
\newblock In A.~J.~M. Spencer, editor, {\em Continuum Theory of the Mechanics
  of Fibre-Reinforced Composites}, pages 1--32. Springer, 1984.

\bibitem{murphy2013transversely}
J.~G. Murphy.
\newblock Transversely isotropic biological, soft tissue must be modelled using
  both anisotropic invariants.
\newblock {\em European Journal of Mechanics A/Solids}, 42:90--96, 2013.

\bibitem{steinmann1997unifying}
P.~Steinmann and E.~Stein.
\newblock A unifying treatise of variational principles for two types of
  micropolar continua.
\newblock {\em Acta Mechanica}, 121(1):215--232, 1997.

\bibitem{pietraszkiewicz2009natural}
W.~Pietraszkiewicz and V.~A. Eremeyev.
\newblock On natural strain measures of the non-linear micropolar continuum.
\newblock {\em International Journal of Solids and Structures},
  46(3-4):774--787, 2009.

\bibitem{Chen86}
Z.~Chen.
\newblock On the representation of finite rotation in nonlinear field theory of
  continuum mechanics.
\newblock {\em Applied Mathematics and Mechanics}, 7(11):1017--1026, 1986.

\bibitem{bell1986continuum}
J.~F. Bell.
\newblock Continuum plasticity at finite strain for stress paths of arbitrary
  composition and direction.
\newblock {\em Archive for Rational Mechanics and Analysis}, 84:139--170, 1983.

\bibitem{beatty1992deformations}
M.~F. Beatty and M.~A. Hayes.
\newblock Deformations of an elastic, internally constrained material. {P}art
  1: {H}omogeneous deformations.
\newblock {\em Journal of Elasticity}, 29(1):1--84, 1992.

\end{thebibliography}

\end{document}